\documentclass[11pt]{article}

\usepackage[a4paper,margin=1in]{geometry}
\usepackage{graphicx}
\usepackage{amsmath,amssymb,bm}
\usepackage{hyperref}
\usepackage{cite}
\usepackage{setspace}

\begin{document}
\title{Limits of Rank Recovery in Bilinear Observation Problems}

\author{Seungbeom Choi \\
Department of Physics, Kyunghee University\\
Graduate Program: Quantum Information Science\\
}

\date{}
\maketitle

\begin{abstract}
Bilinear observation problems arise in many physical and information-theoretic settings, where observables and states enter multiplicatively. Rank-based diagnostics are commonly used in such problems to assess the effective dimensionality accessible to observation, often under the implicit assumption that rank deficiency can be resolved through numerical refinement.

Here we examine this assumption by analyzing the rank and nullity of a bilinear observation operator under systematic tolerance variation. Rather than focusing on a specific reconstruction algorithm, we study the operator directly and identify extended rank plateaus that persist across broad tolerance ranges. These plateaus indicate stable dimensional deficits that are not removed by refinement procedures applied within a fixed problem definition.

To investigate the origin of this behavior, we resolve the nullspace into algebraic sectors defined by the block structure of the variables. The nullspace exhibits a pronounced but non-exclusive concentration in specific sectors, revealing an organized internal structure rather than uniform dimensional loss. Comparing refinement with explicit modification of the problem formulation further shows that rank recovery in the reported setting requires a change in the structure of the observation problem itself.
Here, “problem modification” refers to changes that alter the bilinear observation \emph{structure} (e.g., admissible operator/state families or coupling constraints), in contrast to refinements that preserve the original formulation such as tolerance adjustment and numerical reparameterizations.

Together, these results delineate limits of rank recovery in bilinear observation problems and clarify the distinction between numerical refinement and problem modification in accessing effective dimensional structure.

\end{abstract}

\section{Introduction}
Bilinear observation problems arise in a wide range of physical and information-theoretic settings, where observables and states enter multiplicatively rather than additively. Such structures appear naturally in inverse problems, measurement models, and reconstruction tasks, and they often motivate efforts to recover missing information by increasing resolution, refining numerical procedures, or improving sampling strategies. In many cases, the success of these approaches is assessed through rank-based criteria, which provide an operational measure of the effective dimensionality ... accessible to observation.\cite{GrossCSQST,FlammiaPSIC,ScottTightIC,AhmedBlindDeconv}

A common implicit assumption in this context is that rank deficiency reflects insufficient refinement: if resolution is increased or numerical thresholds are adjusted appropriately, full rank should eventually be recovered. This perspective treats rank loss primarily as a numerical or representational limitation, suggesting that refinement within a fixed problem definition is, in principle, sufficient to restore access to the full space of interest. However, distinguishing between limitations that can be overcome by refinement and those that originate from the structure of the problem itself remains a nontrivial task.\cite{GrossCSQST,FlammiaPSIC}

\paragraph{Physical anchor: tomographic completeness as a bilinear span question.}
A concrete setting where bilinear structure is unavoidable is quantum tomography, where measurement operators and quantum states jointly determine observed data. In such settings, questions of \emph{completeness} or \emph{identifiability} are frequently assessed via rank- or span-based diagnostics applied to a design matrix constructed from sampled measurement operators and states. Our analysis targets this diagnostic layer: we study when rank deficiency is an artifact of numerical representation (e.g., tolerance choice) versus when it reflects a structural restriction of the fixed bilinear observation formulation, thereby clarifying what rank-based “completeness” can and cannot certify in the ... reported regime.\cite{GrossCSQST,FlammiaPSIC,ScottTightIC}
Here we address this distinction by examining rank behavior in a class of bilinear observation problems under systematic tolerance variation. Rather than focusing on a specific algorithm or reconstruction scheme, we analyze the rank and nullity of the associated observation operator directly. This approach reveals extended rank plateaus under tolerance sweeps, indicating stable dimensional deficits that persist across refinement procedures applied within a fixed problem definition.

To probe the origin of these plateaus, we further resolve the nullspace into algebraic sectors defined by the block structure of the variables and contrast refinement with explicit modification of the problem formulation. This analysis exposes a clear operational boundary: while refinement explores numerical representations of a fixed problem, rank recovery in the reported setting requires modification of the problem definition itself. Together, these results identify limits of rank recovery in bilinear observation problems and clarify the roles of refinement and problem modification in accessing the effective dimensional structure of such systems.
\begin{figure}[t]
  \centering
  \includegraphics[width=\linewidth]{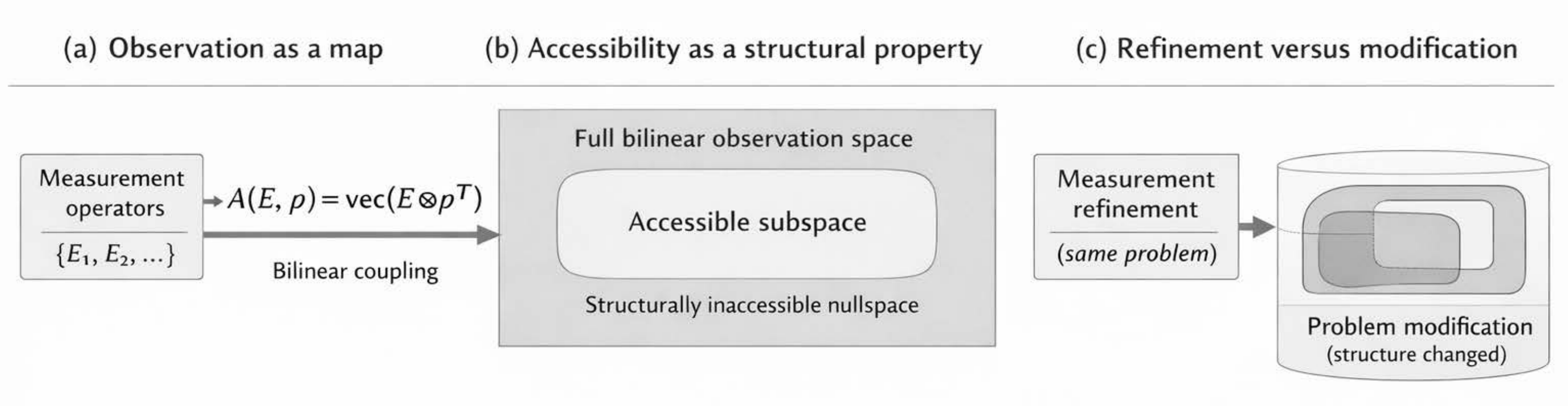}
  \caption{
  \textbf{Conceptual structure of the bilinear observation setting.}
  (a) Observation as a bilinear map from measurement operators and states.
  (b) Accessibility as a structural property, with an accessible subspace and a structurally inaccessible nullspace.
  (c) Distinction between refinement within a fixed formulation and explicit problem modification.
  }
  \label{fig:concept}
\end{figure}

\section{Results}
\subsection{Rank plateaus under tolerance sweeps}

We first examine the rank of the bilinear observation operator under systematic variation of the numerical tolerance used to define nonzero singular values. Across the reported tolerance sweeps, the rank does not vary continuously but instead exhibits extended plateau regions (Fig.~2). \begin{figure}[t]
  \centering
  \includegraphics[width=\linewidth]{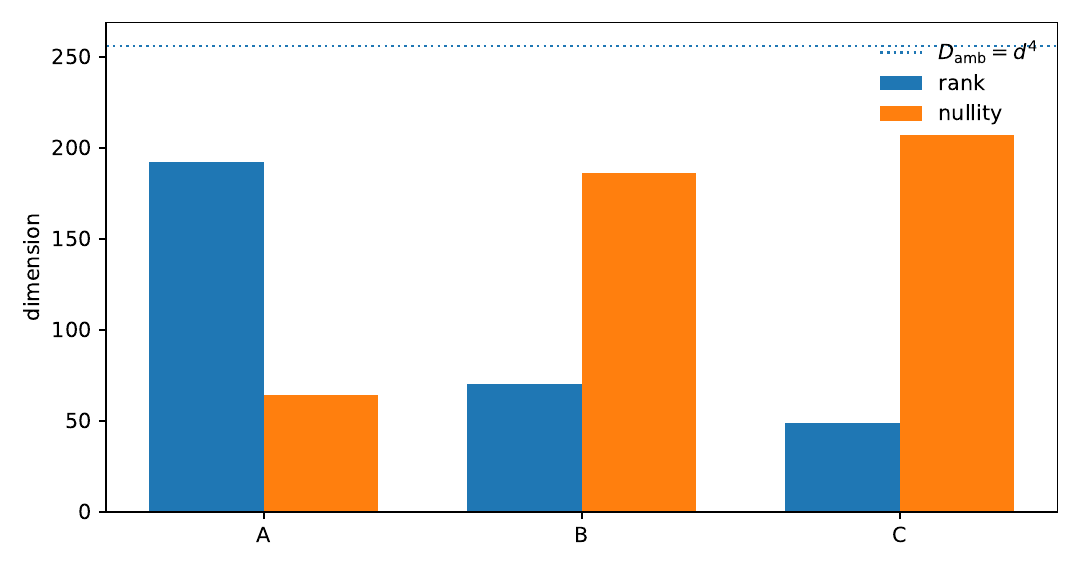}
  \caption{
  \textbf{Rank and nullity under tolerance variation.}
  The rank of the bilinear observation operator is shown as a function of numerical tolerance.
  Extended plateau regions indicate stable rank values that persist across refinement within a fixed problem formulation.
  Nullity is shown as a derived quantity, defined as the difference between the ambient dimension and the measured rank.
  }
  \label{fig:rank-plateau}
\end{figure}
We define a plateau as any maximal consecutive subset of the reported tolerance grid on which $\mathrm{rank}_{\tau}(A)$ is constant; no interpolation or extrapolation beyond the reported grid is used.

 Within each plateau, the rank remains stable despite changes in tolerance, indicating that the observed dimensional reduction is not a transient numerical effect.

All fixed-rank/fixed-nullity statements in this work are restricted to the explicitly reported plateau intervals on the stated tolerance grid; we do not claim invariance outside the reported tolerance range.

The persistence of these plateaus across the reported tolerance ranges suggests that refinement through numerical threshold adjustment alone does not expand the accessible dimensionality of the observation operator. \begin{figure}[t]
  \centering
  \includegraphics[width=\linewidth]{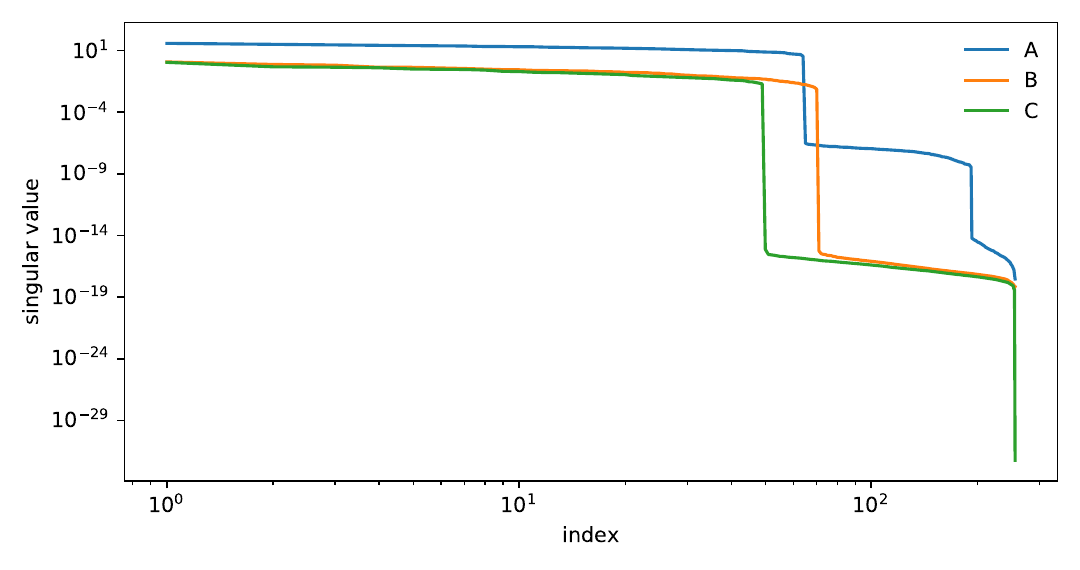}
  \caption{
  \textbf{Robustness of rank plateaus across configurations.}
  Rank-versus-tolerance profiles are shown for multiple experimental configurations.
  The persistence of plateau structure demonstrates that the observed rank behavior is reproducible within the reported settings.
  }
  \label{fig:robustness}
\end{figure}
Rather than converging smoothly toward full rank, the system maintains stable rank values over broad intervals of tolerance.

\subsection{Nullity implied by rank plateaus}

To complement the rank analysis, we report the nullity as a \emph{derived} quantity defined by
\[
\mathrm{nullity}_{\tau}(A) := D_{\mathrm{amb}} - \mathrm{rank}_{\tau}(A).
\]
where $D_{\mathrm{amb}}$ is the ambient dimension of the design matrix under the fixed problem definition. Because nullity is defined directly from rank, any plateau structure observed in rank-versus-tolerance profiles (Figs.~2--3) induces a corresponding plateau structure in nullity over the same tolerance intervals.

Accordingly, throughout this work nullity is used only as a bookkeeping diagnostic that makes explicit the dimensional deficit associated with each observed rank plateau, without introducing an additional measurement or an independent plot-based claim.

\subsection{Sector localization of the nullspace}

To investigate the structural origin of the observed nullity, we resolve the nullspace into algebraic sectors defined by the block structure of the variables. For each tolerance value, we compute the fraction of the nullspace weight associated with each sector (Fig.~4).\begin{figure}[t]
  \centering
  \includegraphics[width=\linewidth]{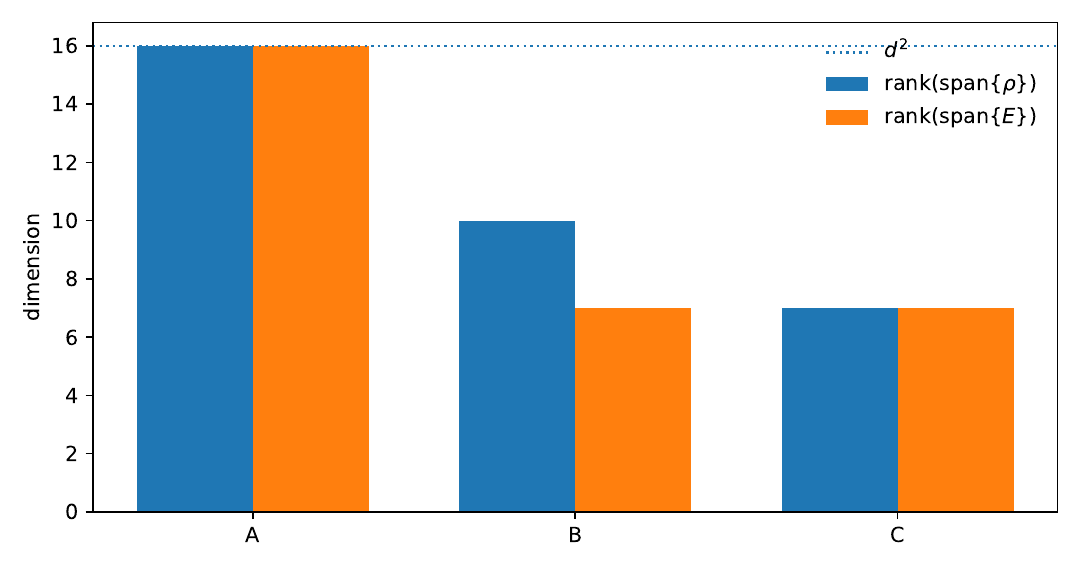}
  \caption{
  \textbf{Sector localization of the nullspace.}
  The nullspace is resolved into algebraic sectors defined by the block structure of the variables.
  A pronounced but non-exclusive concentration is observed in the block-off-diagonal sector, indicating organized nullspace structure.
  }
  \label{fig:sector}
\end{figure}

This sector-resolved analysis reveals a pronounced localization pattern. In the reported runs, a dominant fraction of the nullspace weight is concentrated in the block-off-diagonal sector (see Fig.~4 and Supplementary Note~2 for the reported value at $\tau=1e-12$).
 The localization is therefore strong but not exclusive, indicating that the nullspace structure is organized rather than uniformly distributed.

\subsection{Refinement versus problem modification}

Finally, we compare refinement procedures applied within a fixed problem definition to explicit modification of the problem formulation. Refinement, including tolerance variation and related numerical adjustments, preserves the rank plateaus identified above (Fig.~5). No recovery beyond the reported plateaus is obtained through refinement alone.\begin{figure}[t]
  \centering
  \includegraphics[width=\linewidth]{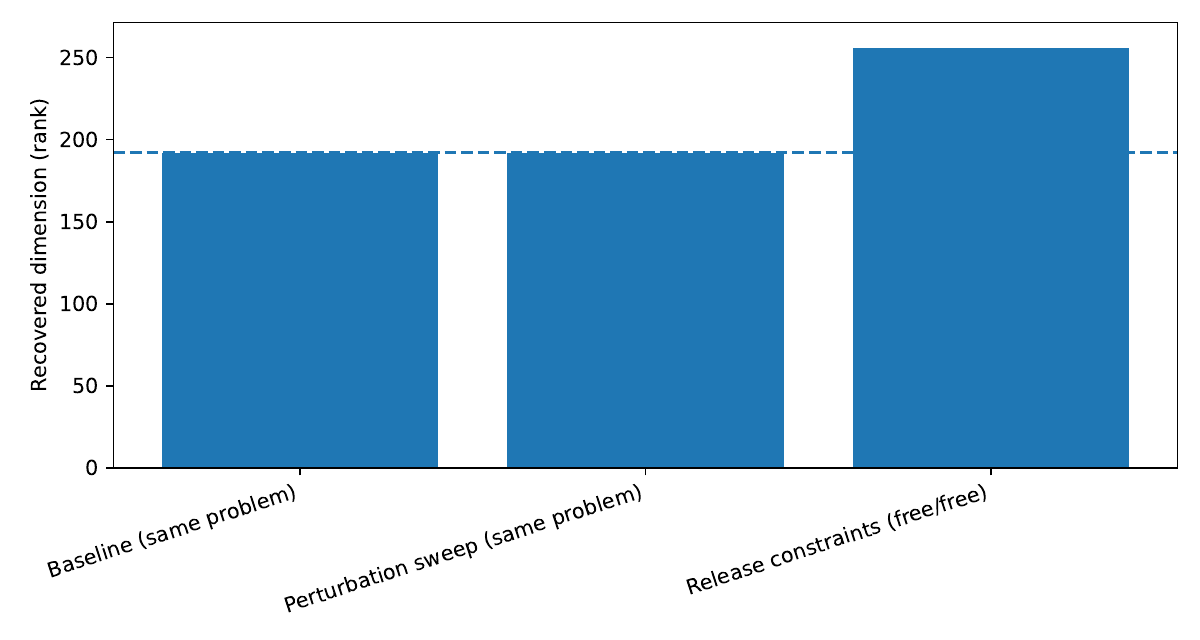}
  \caption{
  \textbf{Refinement versus problem modification.}
  Refinement procedures preserve the observed rank plateaus within a fixed problem formulation.
  In contrast, explicit modification of the observation structure enables recovery of the maximal rank in the reported setting.
  }
  \label{fig:modification}
\end{figure}

In contrast, modifying the problem formulation leads to qualitatively different behavior. Under problem modification, the maximal rank in the ambient setting is recovered in the reported cases (Fig.~5), demonstrating that rank recovery requires a change in the structure of the observation problem rather than further refinement of its numerical representation. This comparison establishes an operational distinction between refinement and problem modification in bilinear observation problems. Importantly, the modification changes the admissible operator/state families (and thus the observation structure), rather than improving the numerical representation of the same fixed formulation.

\section{Methods}
We consider a bilinear observation problem in which an observable $E$ and a state $\rho$ enter multiplicatively through a linear map. For fixed choices of $E$ and $\rho$, this map is represented by a design matrix whose rank and nullity characterize the effective dimensionality accessible to observation. Throughout the analysis, the ambient dimension is fixed, and rank and nullity are defined with respect to this representation.
\subsection*{Bilinear observation operator and design matrix}

For each sampled pair $(E_i,\rho_j)$ we form the bilinear feature vector
$\varphi(E_i,\rho_j)=\mathrm{vec}(E_i\otimes \rho_j^{\mathsf T})\in\mathbb{C}^{d^4}$.
We then build the complex design matrix $A\in\mathbb{C}^{(n_E n_\rho)\times d^4}$ whose rows are
$\varphi(E_i,\rho_j)^{\mathsf T}$ over all $(i,j)$.
We consider three configurations. For Config A, $(n_E,n_\rho)=(16,16)$, yielding $A\in\mathbb{C}^{256\times256}$.
For Configs B and C, $(n_E,n_\rho)=(20,20)$, yielding $A\in\mathbb{C}^{400\times256}$.
In all cases, the ambient dimension is $D_{\mathrm{amb}}=d^4=256$.

\paragraph{Complex convention versus realification (sanity checks only).}
All ranks and nullities reported in Figs.~2--5 use the complex design matrix
$A\in\mathbb{C}^{(n_E n_\rho)\times 256}$ 
(i.e.\ $256\times256$ for A and $400\times256$ for B,C).
In addition, we sometimes consider the standard realification
$A_{\mathbb{R}}=\begin{pmatrix}\mathrm{Re}\,A & -\mathrm{Im}\,A\\ \mathrm{Im}\,A & \mathrm{Re}\,A\end{pmatrix}\in\mathbb{R}^{512\times512}$
as a numerical sanity check (e.g., to verify consistency of input spans).
These realified ranks are \emph{not} used for manuscript figures and are separated in the released package as legacy sanity artifacts.

\subsection*{Rank and nullity evaluation}

The rank of the design matrix is evaluated numerically using a tolerance-based criterion on its singular values. Singular values exceeding the specified tolerance are treated as nonzero. We use a relative threshold rule, counting $\sigma_k$ as nonzero only when $\sigma_k>\tau\cdot\sigma_{\max}$. Because the threshold is relative ($\sigma_k>\tau\sigma_{\max}$), $\mathrm{rank}_{\tau}(A)$ is invariant under global rescaling $A\mapsto cA$ for any nonzero scalar $c$.

The nullity is defined as the difference between the ambient dimension and the measured rank. Rank and nullity are reported only for the tolerance values explicitly included in the reported sweeps.

\subsection*{Tolerance sweeps}

Tolerance sweeps are performed by varying the numerical threshold used to determine nonzero singular values while keeping the underlying bilinear problem definition fixed. For each tolerance value in the reported sweep, rank and nullity are computed independently. Plateau behavior is identified directly from the resulting rank and nullity profiles without extrapolation beyond the reported ranges.

\subsection*{Sector decomposition}

To analyze the internal structure of the nullspace, the design matrix is decomposed into algebraic sectors defined by the block structure of the variables. For each tolerance value, the nullspace is projected onto these sectors, and the fraction of the total nullspace weight associated with each sector is computed. Sector fractions are normalized to sum to unity and are used to characterize dominant localization patterns.
\subsection*{Sector-resolved nullspace weights}

Let $\mathcal{N}(A)$ denote the numerical nullspace at tolerance $\tau$, and let
$\{n_\ell\}_{\ell=1}^{m(\tau)}$ be an orthonormal basis of $\mathcal{N}(A)$ returned by the SVD-based computation.
We define a set of linear sector projectors $\{P_s\}$ associated with the reported block structure
(e.g., block-diagonal and block-off-diagonal components in the fixed basis used to define the sectors).
The sector basis is fixed once for all sweeps/configurations and is not adapted to the computed singular vectors.
The sector weight is computed as
\[
 w_s(\tau) :=
 \frac{\sum_{\ell=1}^{m(\tau)} \lVert P_s n_\ell \rVert_2^2}
 {\sum_{\ell=1}^{m(\tau)} \lVert n_\ell \rVert_2^2},
 \qquad
 \sum_s w_s(\tau) = 1.
\]
Because the basis $\{n_\ell\}$ is orthonormal and the weights are computed via squared norms summed over the full nullspace,
$w_s(\tau)$ is invariant to the particular orthonormal basis choice of the nullspace.

\subsection*{Refinement and problem modification}

Two classes of procedures are considered. Refinement consists of adjustments that preserve the original problem definition, including tolerance variation and related numerical modifications. Problem modification refers to explicit changes in the formulation of the bilinear observation problem. Rank behavior under refinement and under problem modification is evaluated separately and compared to distinguish their effects.

\subsection*{Scope and limitations}

All reported results are restricted to the configurations, tolerance ranges, and sector decompositions explicitly described above. No claims are made regarding behavior outside these reported settings.

\section{Discussion}
The results reported here clarify an important limitation in the interpretation of rank-based diagnostics in bilinear observation problems. Extended rank and nullity plateaus persist under systematic tolerance variation, indicating that refinement of numerical criteria alone does not necessarily increase the effective dimensionality accessible to observation. In the reported setting, rank deficiency is therefore not a transient artifact of threshold choice but a stable feature of the fixed problem formulation.\cite{AhmedBlindDeconv}

By resolving the nullspace into algebraic sectors, we further show that this dimensional deficit is structurally organized rather than uniformly distributed. The observed concentration of nullspace weight in specific sectors demonstrates that rank loss is associated with identifiable components of the bilinear structure. At the same time, the localization is not exclusive, underscoring that the nullspace retains a distributed character rather than collapsing onto a single mode.

A key distinction emerging from this analysis is the operational difference between refinement and problem modification. Refinement procedures, including tolerance sweeps and related numerical adjustments, explore different representations of the same bilinear problem and leave the observed rank plateaus unchanged. In contrast, modifying the problem formulation alters the structure of the observation operator and enables rank recovery in the reported cases. This separation provides a concrete criterion for diagnosing whether rank deficiency reflects a numerical limitation or a structural constraint.

These findings provide a concrete, fully specified reference regime in which rank-based diagnostics exhibit extended plateaus under refinement, and may be informative when interpreting analogous bilinear diagnostics in related inverse or reconstruction settings. Rank-based measures are often used to assess identifiability or completeness, implicitly assuming that sufficient refinement will restore access to the full space. The present results demonstrate that such assumptions must be qualified: within a fixed problem definition, refinement may be fundamentally insufficient. The key point is that not all interventions are comparable: refinements operate within an equivalence class of representations of the \emph{same} bilinear observation problem, whereas structural modifications move the problem to a different equivalence class with different accessible span. Rank recovery under modification therefore does not imply that refinement would eventually succeed; it certifies that the original formulation itself constrains the observable span in the reported regime.

We emphasize that the conclusions are restricted to the configurations, tolerance ranges, and sector decompositions explicitly reported here. The results do not establish universal limits for all bilinear observation problems, nor do they preclude alternative formulations that avoid the observed constraints. Rather, they delineate a concrete regime in which rank recovery is limited by structure rather than by numerical resolution, providing a reference point for future theoretical and experimental investigations.

\section*{Figure legends}

\noindent\textbf{Figure 1 | Conceptual structure of the bilinear observation setting.}
Schematic overview of the bilinear observation operator and the distinction between refinement within a fixed formulation and modification of the formulation. The panel previews the types of robustness and structure diagnostics quantified in Figures 2--5 without asserting any numerical outcome.

\vspace{0.6em}
\noindent\textbf{Figure 2 | Rank of the bilinear observation operator under tolerance variation.}
The accessible dimension (rank) of the bilinear observation operator is shown as a function of the numerical tolerance used to define nonzero singular values. Extended plateau regions are observed, indicating stable rank values that persist across broad tolerance ranges. This behavior shows that varying the tolerance (refinement within a fixed formulation) does not continuously increase the accessible dimensionality in the reported setting.

\vspace{0.6em}
\noindent\textbf{Figure 3 | Robustness of rank plateaus across experimental configurations.}
Rank-versus-tolerance profiles are overlaid for multiple experimental configurations (as listed in the legend). The persistence of plateau structure across configurations supports that the observed rank behavior is not tied to a single run but is reproducible across the reported settings. Each curve is computed from the same tolerance rule applied to the corresponding singular-value spectrum.
Reproducible here means: within the explicitly listed configurations (Config A/B/C) on the stated tolerance grid, using the provided NPZ inputs.

\vspace{0.6em}
\noindent\textbf{Figure 4 | Sector localization of the nullspace.}
The fraction of the nullspace norm associated with each algebraic sector is shown. A pronounced but non-exclusive concentration is observed in the block-off-diagonal sector, while the remaining weight is distributed among other sectors. This localization indicates an organized nullspace structure rather than uniform dimensional loss.

\vspace{0.6em}
\noindent\textbf{Figure 5 | Refinement versus problem modification.}
Recovered rank values are compared for refinement procedures applied within the same problem formulation and for explicit modification of the observation problem. Refinement preserves the observed plateau behavior, whereas problem modification yields higher recovered rank in the reported setting. The comparison distinguishes numerical refinement from structural change in accessing dimensionality.

\end{document}